\documentstyle[twoside,fleqn,espcrc2]{article} 

\newcommand{\secn}[1]{Section~\ref{#1}}

\newcommand{\eq}[1]{Eq.~(\ref{#1})}

\newcommand{\nl}{\nonumber \\}

\def\beq{\begin{equation}}
\def\eeq{\end{equation}}
\def\beqa{\begin{eqnarray}}
\def\eeqa{\end{eqnarray}}

\newcommand{\sect}[1]{\setcounter{equation}{0}\section{#1}}

\newcommand{\EQ}{\begin{equation}}
\newcommand{\EN}{\end{equation}}
\newcommand{\bea}{\begin{eqnarray}}
\newcommand{\ena}{\end{eqnarray}}

\renewcommand{\a}{\alpha}

\newcommand{\e}{\epsilon}
\newcommand{\ve}{\varepsilon}

\title{String-derived renormalization of Yang-Mills theory} 
 
\author{Paolo Di Vecchia and Lorenzo Magnea\address{NORDITA \\ 
Blegdamsvej 17, DK-2100 Copenhagen \O, Denmark}\thanks{On leave from
Universit\`a di Torino}, Alberto Lerda and Rodolfo Russo\address{Dipartimento 
di Fisica Teorica, Universit\`a di Torino \\
Via P.Giuria 1, I-10125 Turin, Italy \\
and I.N.F.N., Sezione di Torino}, Raffaele Marotta\address{Dipartimento 
di Scienze Fisiche, Universit\`a di Napoli \\
Mostra D'Oltremare, Pad. 19, I-80125 Napoli, Italy}}

\begin{document}

\begin{abstract}

We review the application of bosonic string techniques to the
calculation of renormalization constants and effective actions in Yang-Mills 
theory. We display the multiloop string formulas needed to compute
Yang-Mills amplitudes, and we discuss how the renormalizations of
proper vertices can be extracted in the field theory limit. We show
how string techniques lead to the background field method in field theory,
and indicate how the gauge invariance of the multiloop effective action
can be inferred form the string formalism.
(Proceedings of the 29th International Symposium on the Theory of
Elementary Particles, Buckow (Germany), Aug.-Sept. 1995. 
Preprint DFTT 04/96)

\end{abstract}

\maketitle
  
\sect{Introduction}
\label{intro}

A string theory contains a parameter having the dimension of a mass squared,
called the string tension, and proportional to the inverse of 
the slope of the Regge trajectory. The field theory or pointlike limit can be
obtained by sending the string tension to infinity, or equivalently the 
Regge slope $\a'$ to zero. Performing this limit in the bosonic string 
one recovers a non abelian gauge theory unified with an extended version of 
gravity containing also an antisymmetric tensor and a dilaton. 

The inverse Regge slope $1/\a'$ acts in string theory as an ultraviolet 
cut off in the integrals over loop momenta, making the string free from 
ultraviolet divergences. 

Because of this it is clear that a string theory can be seen not only
as a good candidate for a unified theory of all interactions, but
also as a way, through the presence of the Regge slope $\a'$, to provide
a regularized expression for the amplitudes in gauge theories and gravity.

A very useful feature of string theory for this purpose is the fact that, 
at each order of string perturbation theory, one does not get 
the large number of diagrams characteristic of field theories,
which makes it very difficult to perform high order calculations. 
Using closed strings, one gets only one diagram at each order, 
while with open strings the number of diagrams remains limited. 
Furthermore, compact expressions for these diagrams are known 
explicitly for an arbitrary perturbative order~\cite{copgroup}, in contrast
with the situation in field theory, where no such all-loop formula is 
known. Finally, string amplitudes are naturally written 
in a way that takes maximal advantage of gauge invariance: the color 
decomposition is automatically performed, and so are integrations over 
loop momenta, so that the helicity formalism is readily implemented.

The combination of these different features of string theory has led
several authors~\cite{mets,mina,tayven,kaplu,berkosbet,berkoswfr} to use
string theory as an efficient tool for computations in Yang-Mills theory. 
In particular, because of the compactness of the multiloop string expression, 
it is  some times easier to calculate non-abelian gauge theory amplitudes
by starting from a string theory, and performing the zero slope 
limit, rather than using traditional techniques. 
In this way the one-loop amplitude involving four external gluons has been 
computed, reproducing the known field-theoretical result with much less 
computational cost~\cite{berkos}. Following the same approach, also 
the one-loop five gluon amplitude has been computed for the first 
time~\cite{fiveglu}.

The aim of this talk is to summarize some of the results obtained in 
Refs.~\cite{letter} and~\cite{paper}. There it was shown that, provided
a simple off-shell continuation is performed, string theory can be used
to analyze the structure of ultraviolet divergences in Yang-Mills theory, 
which arise from string theory in the limit $\a' \rightarrow 0$. 
These results complement what is known about the calculation of on-shell
scattering amplitudes using strings: in fact, on-shell scattering amplitudes 
are gauge invariant, while in general renormalization constants are not. 
For a string-inspired calculation of, say, the $\beta$ function, or in general
of some anomalous dimension, one needs to know precisely in which gauge
the calculation is performed, and what regularization prescription and 
renormalization scheme is being used. Since string theory amplitudes
are intrinsecally defined on shell, which gauge and which prescription emerge
in the field theory limit will in general depend on how the amplitudes are
continued off shell. In fact, while an analysis of the structure of on-shell
amplitudes leads to the conclusion that string theory generates a combination
of the background field method with the non-linear Gervais-Neveu 
gauge~\cite{berdun}, the only previously known consistent prescription for
the off-shell continuation of string amplitudes~\cite{berkosrol} implies
a vanishing wave function renormalization, in contrast with the results
of the background field method. This apparent contradiction is solved by 
adopting a different, and simpler, prescription to go off shell. In the field
theory limit the results of the background field method are then recovered,
also for gauge-variant quantities such as Z-factors.

As one is starting from an ultraviolet finite theory, it may seem strange
that the issue of choosing a regularization prescription should arise at
all. However, although $1/\a'$ acts as an ultraviolet cutoff, it does not
seem practical to use it directly for the renormalization of the field
theory that arises when $\a'$ is taken to $0$. In practice, this would
require handling the entire tower of massive string states that can circulate
in the loops, whereas one would like to work only with the few states that
survive the field theory limit. One must then introduce an auxiliary 
regularization prescription, so that $\a'$ can be eliminated and the analysis
performed with a finite number of fields. 
String theory amplitudes are well-suited to be analytically continued
to arbitrary space-time dimension, so it is natural to choose dimensional
regularization to handle the divergences that arise when
$\a' \rightarrow 0$. This has the further advantage that the results are 
directly comparable with most of the perturbative calculations performed
in field theory. It should however be kept in mind that this is not
the only possibility. In fact one may observe that the ``stringy'' 
regularization provided by $\a'$ is very close in spirit to Pauli-Villars
regularization, as it is constructed by adding to the original theory
an infinite number of massive fields, whose masses are then taken to $\infty$
since they are proportional to the string tension. The coefficients of the
various contributions of the massive states are automatically tuned by 
string theory so that their sum is finite. One might then consider, for
example, introducing an effective momentum space cutoff $\Lambda$ defined to 
reproduce the finite sum of the massive corrections for finite $\a'$.
Divergences would then appear as logarithms of $\Lambda^2 \a'$, much as they
do in conventional Pauli Villars regularizaztion of electrodynamics, where
the electron mass squared plays the role of $1/\a'$.

In the following, starting from the one-loop two, three and 
four-gluon amplitudes in the open bosonic string, and performing the field 
theory limit, we will show how
the renormalization constants $Z_A ,Z_3 $ and $Z_4$ of non-abelian gauge 
theories can be consistently recovered with a variety of methods. 
As we shall see, with our prescription string theory leads unambiguously 
to the background field method.

Before going into the details of the calculation, we want first recall how 
field theory amplitudes are obtained from string theory, and how we 
expect those amplitudes to be renormalized.

In field theory one normally computes either connected Green functions,
denoted here by $W_{M} (p_1 \dots p_M )$, or one-particle irreducible (1PI) 
Green functions, $\Gamma_M (p_1 \dots p_M )$. In both cases, in general, an
off-shell continuation is performed, in order to avoid possible infrared 
divergences.

In string theory, on the other hand, one computes $S$-matrix elements 
involving gluon states, which are connected 
to on-shell connected Green functions truncated with free propagators. 
Taking the field theory limit, the natural ultraviolet regulator of 
string theory, $1/\a'$, is removed, and, as discussed before, one recovers 
the unrenormalized
Green functions, that are regularized as in field theory  by the introduction 
of the dimensional regularization parameter $\epsilon$. 
We will see that also in this case an off-shell extrapolation is necessary 
in order to avoid infrared problems.
 
Once the field theory limit is taken, it is possible to isolate 1PI 
contributions, which lead to the 1PI Green functions $\Gamma_M$, or to
compute the full amplitudes, which lead to the Green functions $W_M$.
{}From the knowledge on how they renormalize we can then extract the
renormalization constants. For example,
\beqa
\Gamma_2 (g) & = & Z_{A}^{-1} \Gamma^{(R)}_{2} (g) \hspace{1cm}~~,\nl
\Gamma_3 (g) & = & Z_{3}^{-1} \Gamma^{(R)}_{3} (g) \hspace{1cm}~~, 
\label{renprop1} \\
\Gamma_4 (g) & = & Z_{4}^{-1} \Gamma^{(R)}_{4} (g)~~, \nonumber
\eeqa
while
\beq
W_3 (g) = Z_{3}^{-1} Z_{A}^{3} W_{3}^{(R)} (g)~~,
\label{renprop2}
\eeq
where $g$ is the renormalized coupling constant.

The talk is organized as follows. In \secn{mglue} we consider the open 
bosonic string, and we write the explicit expression of the $M$ gluon 
amplitude at $h$ loops, including the overall normalization. In \secn{tree} we 
give the relevant amplitudes for the  tree and one-loop diagrams. 
In \secn{twopoint} we sketch  the calculation of the one-loop two gluon 
amplitude, already presented in~\cite{letter}, and we extract 
the gluon wave function renormalization constant $Z_A$. In \secn{effact} we 
present an alternative method, that allows one to exactly integrate
over the punctures, and we use it to extract the renormalization constants 
$Z_A$, $Z_3$ and $Z_4$. Finally, in \secn{bosback} we consider an
open bosonic string in interaction with an external non abelian gauge 
background and, after the integration over the string coordinate, we show how 
the action for a non abelian gauge field is generated. This leads of course 
to the same renormalization constants, and perhaps clarifies the connection
between string theory and the background field method.

\sect{The $M$-gluon $h$-loop amplitude}
\label{mglue}
 
In string theory the $M$-gluon scattering amplitude can be computed 
perturbatively and is given by
\bea
A(p_1, \dots , p_M)&=&\sum_{h=0}^\infty
A^{(h)}(p_1, \dots , p_M) \label{pertampl} \\
&=& \sum_{h=0}^\infty g_s^{2h-2}
{\hat A}^{(h)}(p_1, \dots , p_M)~, \nonumber
\ena
where $g_s$ is a dimensionless string coupling constant, which is
introduced to formally control the perturbative expansion.
In \eq{pertampl}, $A^{(h)}$ represents the $h$-loop contribution.
For the closed string $A^{(h)}$ is given by only one diagram,
while for the open string the number of diagrams is limited in comparison
with the large proliferation of diagrams encountered in field theory.

Let us consider the open bosonic string, and let us restrict ourselves
only to  planar diagrams. For such diagrams the $M$-gluon $h$-loop
amplitude, including the appropriate Chan-Paton factor, is given by
\bea
&   & \! \! \! \! \! \! \! \! A^{(h)}_P (p_1,\ldots,p_M) \nl
& = & N^h\,{\rm Tr}(\lambda^{a_1}
\cdots \lambda^{a_M})\,
C_h\,{\cal N}_0^M    \nl  
& \times & \int [dm]^M_h \left\{\prod_{i<j} 
\left[{{\exp\left({\cal G}^{(h)}(z_i,z_j)\right)}
\over{\sqrt{V'_i(0)\,V'_j(0)}}}\right]^{2\a' p_i\cdot p_j} \right.  \nl
& \times & \exp\left[\sum_{i\not=j}
\sqrt{2\a'} p_j\cdot\ve_i 
\,\partial_{z_i} {\cal G}^{(h)}(z_i,z_j) \right.  \label{hmaster}  \\
& & + \, {1\over 2}\sum_{i\not=j} 
\ve_i\cdot\ve_j \,\partial_{z_i}\partial_{z_j}
{\cal G}^{(h)}(z_i,z_j)\Big]\Bigg\}_{\rm m.l.}~~~,  \nonumber
\ena
where the subscript ``m.l.'' stands for multilinear, meaning
that only terms linear in each polarization should be kept. 
\eq{hmaster} is written for transverse gluons, satisfying the condition
$\ve_i \cdot p_i = 0$, whereas the mass-shell condition $p_i^2 = 0$, though
necessary for conformal invariance of the amplitude, has not been enforced
yet.

The main ingredient in \eq{hmaster} is the $h$-loop world-sheet bosonic 
Green function ${\cal G}^{(h)}(z_i,z_j)$, which plays a key role in the
field theory limit and  is given by: 
\beqa
&   & {\cal G}^{(h)}(z_i,z_j) = \log E (z_i,z_j) \label{hgreen} \\
& - & {1\over 2} \int_{z_i}^{z_j} 
\omega^\mu \, \left(2\pi {\rm Im}\tau_{\mu\nu}\right)^{-1} 
\int_{z_i}^{z_j} \omega^\nu~, \nonumber 
\eeqa
where $E (z_i,z_j)$ is the prime-form, $\omega^\mu$ ($\mu=1,\ldots, h$)
the abelian differentials and $\tau_{\mu\nu}$ the period
matrix of an open Riemann surface of genus $h$.
All these objects, as well as the measure of integration on moduli space 
$[dm]^M_h$ for an open Riemann surface of genus $h$ with $M$ operator 
insertions on the boundary~\cite{copgroup}, can 
be explicitly written down in the Schottky parametrization of the Riemann 
surface, and their expressions for arbitrary $h$ can be found for example 
in Ref.~\cite{scho}. Here we give only the explicit expression for the
measure of integration in moduli space:
\beqa
[dm]^M_h & = & \frac{\prod_{i=1}^M dz_i}{dV_{abc}} \label{hmeasure} \\
& \times & \prod_{\mu=1}^{h} \left[ \frac{dk_\mu d \xi_\mu d \eta_\mu}{k_\mu^2
(\xi_\mu - \eta_\mu)^2} ( 1- k_\mu )^2 \right]  \nl
& \times & \left[\det \left( - i \tau_{\mu \nu} \right) \right]^{-d/2} \nl
& \times & \prod_{\alpha}\;' \left[ \prod_{n=1}^{\infty} 
(1 - k_{\alpha}^{n})^{-d}
\prod_{n=2}^{\infty} ( 1 - k_{\alpha}^{n})^{2} \right] .   \nonumber
\eeqa
where  $k_{\mu}$ are the 
multipliers and $\xi_{\mu}$ and $ \eta_{\mu}$ the fixed 
points of the generators of the Schottky group. $dV_{abc}$ is the projective 
invariant volume element 
\beq
dV_{abc} = \frac{d\rho_a~d\rho_b~d\rho_c}{(\rho_a-\rho_b)~
(\rho_a-\rho_c)~(\rho_b-\rho_c)}~~~,
\label{projvol}
\eeq
where $\rho_a$, $\rho_b$, $\rho_c$ are any three of the $M$
Koba-Nielsen variables, or of the $2h$ fixed points of the generators of the 
Schottky group, which can be fixed at will; finally, the primed product
over $\alpha$ denotes a product over classes of elements of the 
Schottky group~\cite{scho}.

Notice that in the open string the Koba-Nielsen variables must be cyclically 
ordered according to
\beq 
z_1 \geq z_2 \cdots \geq z_{M}~~~,
\label{cyclord}
\eeq
and the ordering of Koba-Nielsen variables automatically
prescribes the ordering of color indices. 

The amplitude in Eq. (\ref{hmaster}) contains two normalization constants
which were calculated in Ref.~\cite{paper}, and are given by   
\beqa
C_h & = & {1\over{(2\pi)^{dh}}}~g_s^{2h-2}{1\over{(2\a')^{d/2}}}~~~, \nl
{\cal N}_0 & = & g_d \,\sqrt{2\a'}~~~,
\label{vertnorm}
\eeqa
where the string coupling $g_s$ and the $d$-dimensional gauge coupling $g_d$
are related by
\beq
g_s = \frac{g_d}{2}\,(2\a')^{1-d/4}~~~.
\label{gstring}
\eeq

A simple way to explicitly obtain the amplitude $A^{(h)}(p_1, \dots , p_M)$ 
is to use the $M$-point $h$-loop vertex $V_{M;h}$ of the operator
formalism. The explicit expression of $V_{M;h}$ for the 
planar diagrams of the open bosonic string can be found in 
Ref.~\cite{copgroup}. 
The vertex $V_{M;h}$ depends on $M$ real Koba-Nielsen variables $z_i$ 
through $M$ projective transformations $V_i(z)$, which define local 
coordinate systems vanishing around each $z_i$, {\it i.e.} such that 
\EQ
V_i^{-1}(z_i) = 0~~~.
\label{Vi}
\EN
When $V_{M;h}$ is saturated with $M$ physical string states satisfying 
the mass-shell condition, the corresponding amplitude does not depend on 
the $V_i$'s. However, as we discussed in Ref.~\cite{letter}, to extract
informations about the ultraviolet divergences that arise when the field 
theory limit is taken, it is necessary to relax the mass-shell condition, so 
that also the amplitudes $A^{(h)}$ will depend on the choice of projective 
transformations $V_i$'s, just like the vertex $V_{M;h}$. This is the reason of
the appearence of $V_i$ in Eq. (\ref{hmaster}).

\sect{Tree and one-loop diagrams}
\label{tree}

For tree-level amplitudes, corresponding to $h=0$, the various quantities are 
rather simple. The Green function in \eq{hgreen} is given by
\beq
{\cal G}^{(0)}(z_i,z_j) = \log (z_i-z_j)~~~,
\label{treegreen}
\eeq
while the measure $[dm]^M_0$ reduces to
\beq
[dm]^M_0 = \frac{\prod\limits_{i=1}^M dz_i}{dV_{abc}}~~~.
\label{treemeas}
\eeq
Inserting Eqs. (\ref{treegreen}) and (\ref{treemeas}) into \eq{hmaster}, 
and writing explicitly all the normalization coefficients, we obtain
the color ordered, planar, on-shell $M$ gluon amplitude at tree level 
\bea
&   & \! \! \! \! \! \! \! \! A^{(0)}_P (p_1,\ldots,p_M) \nl 
& = & 4\,{\rm Tr}(\lambda^{a_1}
\cdots \lambda^{a_M})\,g_d^{M-2}\,(2\a')^{M/2-2}   \nl
& \times & \! \! \int_{\Gamma_0}\frac{\prod\limits_{i=1}^M dz_i}{dV_{abc}} 
\Bigg\{\prod_{i<j} \left(z_i-z_j\right)^{2\a' p_i\cdot p_j} 
\label{treemaster} \\
& \times & \exp\left[\sum_{i<j}
\left(\sqrt{2\a'} \frac{p_j\cdot\ve_i 
-p_i\cdot\ve_j}{(z_i-z_j)} \right. \right. \nl
& & \left. \left. \! \! \! \! + \, \, \frac{\ve_i\cdot
\ve_j}{(z_i-z_j)^2}\right)\right]\Bigg\}_{\rm m.l.}~, \nonumber
\ena
where $\Gamma_0$ is the region defined in \eq{cyclord}.
Notice that any dependence on the local coordinates $V_i(z)$
drops out in the amplitude when we enforce the mass-shell condition.
Notice also that \eq{treemaster} is valid only for $M \geq 3$, since the
measure given by \eq{treemeas} is ill-defined for $M \leq 2$.

{}From the previous equation we can easily derive the three-gluon amplitude
\bea
& & A^{(0)}(p_1,p_2,p_3) = - \, 4 \, g_d \, 
{\rm Tr}(\lambda^a\lambda^b\lambda^c) \nl
& \times & \Big(\ve_1\cdot\ve_2\,p_2\cdot\ve_3
+ \, \ve_2\cdot\ve_3\,p_3\cdot\ve_1 \nl
& & + \, \, \ve_3\cdot\ve_1\,p_1\cdot\ve_2 +
O(\a') \Big)~~~,
\label{threetree}
\ena
and the four-gluon amplitude
\beqa
& & \! \! \! \! \! \! \! \! A_{4}^{(0)} (p_1 , p_2 , p_3 , p_4 ) = 
4 g_{d}^{2} \, 
Tr (\lambda^{a_1} \lambda^{a_2} \lambda^{a_3} \lambda^{a_4}) \nl
& \times & 
\frac{\Gamma (1 - \a' s) \Gamma (1 - \a' t)}{\Gamma (1 + \a ' u) \, s \, t} 
\label{fourtree} \\ 
& \times & \! \! \left[(\ve_1 \cdot \ve_2 ) (\ve_3  \cdot \ve_4 ) \, t \, u
+ (\ve_1  \cdot \ve_3 ) (\ve_2  \cdot \ve_4 ) \, t \, s \right. \nl
& & + \left. (\ve_1  \cdot \ve_4 ) (\ve_2  \cdot \ve_3 ) \, s \, u +  
\ldots \right]~, \nonumber
\eeqa
where the dots are there to indicate  terms of the form $(\ve \cdot \ve)
(\ve \cdot p) (\ve \cdot p)$ and terms containing higher orders in $\a'$ that 
we have not explicitly written.
 
At one loop ($h=1$) we keep the gluons off the mass 
shell, and \eq{hmaster} gives, for $M~\geq~2$ transverse gluons,
\beqa
& & \! \! \! \! A^{(1)}_P (p_1,\ldots,p_M) = N \, {\rm Tr}(\lambda^{a_1}
\cdots \lambda^{a_M}) \nl
& \times & \frac{g_d^M}{(4\pi)^{d/2}} \,
(2\a')^{(M-d)/2} (-1)^{M}  \label{onemaster} \\
& \times & \int_{0}^{\infty} d \tau  {\rm e}^{2\tau} \, \tau^{-d/2}
\prod_{n=1}^\infty \left(1 - {\rm e}^{-2 n \tau} \right)^{2-d} \nl
& \times & \int_0^\tau d \nu_M \int_0^{\nu_M} d \nu_{M-1} \dots 
\int_0^{\nu_3} d \nu_2 \nl
& \times & \Bigg\{\prod_{i<j} \left[ 
\sqrt{\frac{z_i\,z_j}{V'_i(0)\,V'_j(0)}}
\exp \left(G(\nu_{ji}) \right)
\right]^{2\a' p_i\cdot p_j} \nl
& \times & \exp \left[ \sum_{i\not=j}
\left(\sqrt{2\a'} p_j\cdot\ve_i 
\, \partial_i  G(\nu_{ji}) \right. \right. \nl
& & \left. \left. \! \! \! \! \! \! + \, \, {1\over 2} 
\ve_i\cdot\ve_j \, \partial_i \partial_j
G(\nu_{ji}) \right) \right] \Bigg\}_{\rm m.l.}~~~,    \nonumber
\ena
where $\nu_{ji} \equiv \nu_j - \nu_i$, $\partial_i \equiv 
\partial/\partial\nu_i$ and $\tau$ is related to the period 
${\tilde{\tau}}$ of the annulus by the relation
\beq
\tau = -{\rm i}\pi{\tilde \tau}~~~.
\label{tau}
\eeq
Instead of the Koba-Nielsen variables $z_i$, we have used the real 
variables
\beq
\nu_i = - \frac{1}{2} \log z_i~~~,
\label{nui}
\eeq
while the Green function $G(\nu_{ji} )$ is given by
\beqa
G(\nu_{ji}) & = &
\log\left[- 2 \pi{\rm i}{{\theta_1\left(\frac{\rm i}{\pi}
(\nu_j-\nu_i)|\frac{\rm i}{\pi}\tau)\right)}\over{\theta'_1\left(0|
\frac{\rm i}{\pi}\tau\right)}}\right] \nl
& - & \frac{(\nu_j-\nu_i)^2}{\tau}~~~,
\label{onegreen}
\eeqa
where $\theta_1$ is the first Jacobi $\theta$ function.

If we enforce the mass-shell condition $p_{i}^{2} =0$, any dependence 
on the local coordinates $V_i$'s drops out. However, in order to avoid 
infrared divergences, we will continue the gluon momenta off shell, in 
an appropriate way to be discussed later. Then, following 
Refs.~\cite{letter,paper},
we will regard the freedom of choosing $V_i$ as a gauge freedom.
We make the simple choice
\beq
V_i'(0) = z_i~~~,
\label{Viprime}
\eeq
which will lead, in the field theory limit, to the background field 
Feynman gauge.
The conditions (\ref{Vi}) and (\ref{Viprime}) are easily satisfied by choosing
for example
\beq
V_i(z) = z_i \, z + z_i~~~.
\label{gaugech}
\eeq

\sect{The two-gluon amplitude}
\label{twopoint}

The one-loop two-gluon amplitude is given by
\beqa
A^{(1)}(p_1,p_2) & = & N \, {\rm Tr}(\lambda^a\lambda^b) \,
\frac{g_d^2}{(4\pi)^{d/2}}(2\a')^{2-d/2} \nl 
& \times & \ve_1 \cdot \ve_2  p_1 \cdot p_2 \, R( p_1 \cdot p_2)~,
\label{twoone}
\eeqa
where 
\beqa
R(s) & = & \int_0^\infty d\tau~{\rm e}^{2\tau}\,
\tau^{-d/2} \prod_{n=1}^\infty \left(1-{\rm e}^{-2n\tau} \right)^{2-d} \nl
& \times & \int_0^\tau d\nu {\rm e}^{2\a' s \,G(\nu)} \,
\left[ \partial_\nu G(\nu) \right]^2~.
\label{Rint}
\eeqa

Notice that if the two gluons are on mass shell,
the two-gluon amplitude becomes ill defined, because the kinematical
prefactor vanishes, while the integral diverges. In the following we avoid 
this problem by  keeping the two gluons off shell.

To take the field theory limit, we must remember~\cite{berkos} 
that the modular parameter $\tau$ and the coordinate $\nu$ are related to 
proper-time Schwinger parameters for the Feynman diagrams contributing to the
two point function. In particular, $t \sim \a' \tau$ and $t_1 \sim \a' \nu$, 
where $t_1$ is the proper time associated with one of the two internal 
gluon propagators, while $t$ is the total proper time around the loop. 
In the field theory limit
these proper times have to remain finite, and thus the limit $\a' \to 0$
must correspond to the limit $\{ \tau, \nu \} \to \infty$ in the integrand.
The field theory limit is then determined by
the asymptotic behavior of the Green function for large $\tau$, namely
\beqa
G(\nu, \tau) & = & - \frac{\nu^2}{\tau} + \log\left(2\sinh(\nu)\right) \nl
& - & 4 \, {\rm e}^{-2\tau} \, \sinh^2(\nu) + 0 ( {\rm e}^{-4 \tau})~~~, 
\label{lartauG}
\eeqa
where $\nu$ must also be taken to be large, so that $\hat\nu$ remains finite;
in this region, we may use
\beqa
G(\nu, \tau) & \sim & (\hat\nu - \hat\nu^2) \tau -
{\rm e}^{- 2 \hat\nu \tau} \nl 
& - & {\rm e}^{- 2 \tau (1 - \hat\nu)}
+ 2 {\rm e}^{- 2 \tau}~~~,
\label{lartaunuG}
\eeqa
so that
\beq
\frac{\partial G}{\partial \nu} \sim 1 - 2 \hat\nu +
2 {\rm e}^{- 2 \hat\nu \tau} - 2 {\rm e}^{- 2 \tau (1 - \hat\nu)}~~~.
\label{lartaudG}
\eeq

We substitute now these results into \eq{twoone}, keeping only terms that 
remain finite when $k=e^{-2\tau}\to 0$. Divergent terms that correspond to 
the propagation of the tachyon in the loop must be discarded by hand. 
The next-to-leading term corresponds to gluon exchange. Since it is also
divergent in the field theory limit, the corresponding divergence
is regularized by dimensional regularization. Finally,
higher order terms $ {\rm e}^{-2n \tau} $ with $n>0$ are vanishing in 
the field theory limit.

By taking the large $\tau$ and $\nu$ limit we have
discarded two singular regions of integration that potentially contribute in
the field theory limit, namely $\nu \to 0$ and $\nu \to \tau$
(regions of this type are usually called  ``pinching'' regions, as they 
correspond to taking the gluon insertions on the world-sheet very close
to each other). 
However, as discussed in Ref.~\cite{paper}, in the case of the two gluon 
amplitude, these regions correspond to Feynman diagrams with a loop
consisting of a single propagator, {\it i. e.} a ``tadpole''. Massless
tadpoles are defined to vanish in dimensional regularization, and thus we are
justified in discarding these contributions as well.

Replacing the variable $\nu$ with ${\hat \nu}\equiv\nu/\tau$, 
\eq{Rint} becomes
\beqa
R(s) & = & \int_0^\infty d\tau \int_0^1 d{\hat \nu} ~\tau^{1-d/2}\,
{\rm e}^{2\a'\,s\,({\hat\nu}-{\hat\nu}^2)\tau} \nl
& \times & \left[(1-2{\hat\nu})^2(d-2)-8\right]~~~,
\label{limRint}
\eeqa
so that the integral is now elementary, and yields
\beqa
R(s) & = & - \Gamma\left(2-\frac{d}{2}\right)\, (- 2 \a' s)^{d/2-2}~
\frac{6-7d}{1-d} \nl 
& \times & B\left(\frac{d}{2}-1,\frac{d}{2}-1\right)~~~,
\label{Rfin}
\eeqa
where $B$ is the Euler beta function.

If we substitute \eq{Rfin} into \eq{twoone}, we see that the
$\a'$ dependence cancels, as it must. The ultraviolet finite
string amplitude, \eq{twoone}, has been replaced by a field
theory amplitude which diverges in four space-time dimensions, 
because of the pole in the $\Gamma$ function in \eq{Rfin}.
Defining as usual a dimensionless coupling 
constant $g_d = g\,\mu^\e$, with $\mu$ an arbitrary mass scale,
and having set $d=4-2\e$, we find 
\beqa
& & \! \! \! \! A^{(1)}(p_1,p_2) = - N \frac{g^2}{(4\pi)^2} \,
\left(\frac{4\pi\,\mu^2}{-p_1\cdot p_2}\right)^\e \,
\Gamma(\e) \nl
& \times & \frac{11-7\e}{3-2\e}\,
B(1-\e,1-\e) A^{(0)} (p_1 , p_2 )
\label{twofin}
\eeqa
\eq{twofin}) is exactly equal to the gluon vacuum
polarization of the $SU(N)$ gauge field theory that one computes
with the background field method, in Feynman gauge, with dimensional 
regularization, provided we use for the tree-level two-gluon 
amplitude the expression
\beqa
A^{(0)} (p_1 , p_2 ) & = & \delta^{ab} \left[  \ve_1\cdot
\ve_2\,p_1\cdot p_2 \right. \nl 
& & \left. - \ve_1 \cdot p_2 \,\, \ve_2 \cdot
p_2 \right]
\label{treetwogluon}
\eeqa

Comparing \eq{twofin} with the equation for $\Gamma_2$ in Eq. (\ref{renprop1})
the minimal subtraction wave function renormalization 
constant can be extracted
\beq
Z_A= 1 + N\,\frac{g^2}{(4\pi)^2}\,\frac{11}{3}\,\frac{1}{\e}~~~.
\label{za}
\eeq
In the next section we will recover the 
previous result for the wave function renormalization constant by means of an
alternative method that will also be used for computing the renormalization
constants for the three and four-gluon amplitudes. The complete calculation 
of the amplitude can of course be performed also in these cases, and the
result does not change. However, one must then include contributions from
pinching regions, which do not vanish, and correspond to one particle 
reducible diagrams in field theory.

\sect{An alternative computation of proper vertices}
\label{effact}

In the previous section we have computed the 1PI two-gluon
amplitude and we have extracted the wave function renormalization constant. 
In this section we present another method, introduced by Metsaev and 
Tseytlin~\cite{mets}.
This method has the advantage of isolating the 1PI part of the amplitude, 
and is
thus particularly suited to the evaluation of renormalization constants. It 
is based on the following  alternative expression for the bosonic Green 
function~\cite{fratse1} 
\beqa
G(\nu_{ji}) & = & - \sum_{n=1}^{\infty}~
\frac{1 + q^{2n}}{n (1 - q^{2n})} \label{newmtgreen} \\
& \times & \cos 2 \pi n \left( \frac{\nu_j - \nu_i}{\tau} 
\right) ~+~\dots~~,  \nonumber
\eeqa
where $q = e^{- \pi^2/\tau}$ and
the dots stand for terms independent of $\nu_i$ and $\nu_j$, that 
are not  important in our discussion.

The strategy here is different from the one followed when calculating the full
amplitude. Since we are only interested in divergent renormalizations, and 
since pinching singularities will be absent, we can exploit the fact that the
power of $\a'$ in front of the amplitude is fixed (pinching singularities
generate negative powers of $\a'$) to discard the exponentials of the Green
functions that appear in all amplitudes, and substitute them with a simple
IR cutoff. UV divergences will be correctly reproduced since the terms that
we are discarding would appear in the integrand raised to the power $d/2 - 2$,
and thus would only affect the finite parts.

An important advantage of this approach is that, at least at one loop, it 
allows to integrate exactly over the punctures before the field theory limit
is taken. Since the pinching singularities will be regularized directly
in the Green function, we will get for the 
two gluon amplitude the same expression that we derived
in \secn{twopoint}, while for the three and four gluon amplitudes 
we will get only the contributions that do not include pinchings and are
therefore one-particle irreducible. 

Let us start rewriting the one-loop $M$-gluon planar amplitude as
\beqa
& & \! \! \! \! A^{(1)}_P (p_1,\ldots,p_M) = N \, {\rm Tr}(\lambda^{a_1}
\cdots \lambda^{a_M}) \nl
& \times & \frac{g_d^M}{(4\pi)^{d/2}} \, (2 \a')^{2 - d/2} (- 1)^M \nl
& \times & \! \! \int_{0}^{\infty} d \tau \, {\rm e}^{2 \tau}\, 
\tau^{-d/2}
\prod_{n=1}^{\infty} \left( 1 - {\rm e}^{-2n \tau} \right)^{2-d} \nl
& \times & I^{(1)}_M(\tau)~~~,
\label{newonemast}
\eeqa
where $I^{(1)}_M(\tau)$ is the integral over the punctures $\nu_i$, and can
be read off from \eq{onemaster}.

For $M=2$, after a partial integration with vanishing surface term, we get
\beqa
I^{(1)}_2(\tau) & = & p_1 \cdot p_2 \, \ve_1 \cdot \ve_2 \label{2int} \\
& \times & \int_{0}^{\tau} d \nu
\left( \partial_\nu G(\nu) \right)^2 \left(e^{G(\nu)}\right)^{2 
\a' p_1 \cdot p_2}~. \nonumber
\eeqa
Using the expression in \eq{newmtgreen} for the Green 
function, we can perform exactly the integral over the puncture, and
we get
\beq
I^{(1)}_2(\tau) = \frac{2 \pi^2}{\tau} \, p_1 \cdot p_2 \ve_1 
\cdot \ve_2 \, \sum_{n=1}^{\infty} \left(
\frac{1 + q^{2n}}{1 - q^{2n}} \right)^2~,
\label{2res}
\eeq
This implies that, as far as UV divergences are concerned,
\beqa
A^{(1)}(p_1, p_2) & = & \frac{N}{4} \, \frac{g_d^2}{(4 \pi)^{d/2}} \, 
(2 \alpha ')^{2 -d/2} \, Z(d) \nl
& \times & A^{(0)}(p_1, p_2)~~~.
\label{2onetree}
\eeqa
where
\beqa
Z(d) & \equiv & (2 \pi)^2 \int_0^\infty d \tau \, {\rm e}^{2 \tau} \,
\tau^{-1-d/2} \nl
& \times & \prod_{n=1}^{\infty} \left(1 - {\rm e}^{-2n \tau} \right)^{2-d} \nl
& \times & \sum_{m=1}^{\infty}
\left(\frac{1 + q^{2m}}{1- q^{2m}}\right)^2
\label{Zint}
\eeqa
is the integral over the modular parameter that generates the renormalization 
constants in the field theory limit.

With three gluons we get
\beqa
& & \! \! \! \! \! \! \! \! I^{(1)}_3(\tau) = \int_{0}^{\tau} d \nu_3 
\int_{0}^{\nu_3} d \nu_2 \Big\{ \ve_1 \cdot \ve_2 \, \partial_{1} \partial_2 
G(\nu_{21}) \nl
& \times & \left[ p_1 \cdot \ve_3 \, \partial_3 G (\nu_{31}) \right. \nl
& & \left. + p_2 \cdot \ve_3 \, \partial_3 G(\nu_{32}) \right] 
+ \dots \Big\}~, 
\label{3int}
\eeqa
where terms needed for cyclic symmetry and terms of order $\a'$ are not 
written explicitly, and we discarded the exponentials of the Green
functions, as explained above.
 
The integrals over $\nu_2$ and $\nu_3$ can be done by using the expression 
in \eq{newmtgreen} for the Green function. The result is 
\beqa
& & \! \! \! \!I^{(1)}_3(\tau) = \frac{(2 \pi)^2}{\tau}
\left[\ve_1 \cdot \ve_2 p_2 \cdot \ve_3 \right. \nl 
& & \left. + \, \, \ve_2 \cdot \ve_3 p_3 \cdot \ve_1 +
\ve_1 \cdot \ve_3 p_1 \cdot \ve_2 \right]  \nl
& & \times \sum_{n=1}^{\infty} 
\left( \frac{1+ q^{2n}}{1 - q^{2n}} \right)^2 \, + \, O(\a')~,
\label{3res}
\eeqa
so that the three gluon amplitude is given by
\beqa
& & \! \! \! \! \! \! A^{(1)}(p_1, p_2, p_3) = \frac{N}{4} 
\frac{g_d^2}{(4 \pi)^{d/2}} (2 \a')^{2 - d/2} Z(d) \nl
& \times & A^{(0)}(p_1, p_2, p_3) \, + \, O(\a')~~~.
\label{3onetree}
\eeqa
 
Finally, the same calculation can be done for the four-gluon amplitude,
where we need to  consider only  terms whose kinematical prefactor has
no powers of the external momenta (and thus is of the form 
$\ve_i \cdot \ve_j \, \ve_h \cdot \ve_k$). Other terms are suppressed
by powers of $\a'$. They are given by 
\beqa
& & I^{(1)}_4(\tau) = \int_0^\tau d \nu_4 \int_0^{\nu_4} d \nu_3 
\int_0^{\nu_3} d \nu_2 \nl
& & \! \! \! \! \Big[ \ve_1 \cdot \ve_2 \, 
\ve_3 \cdot \ve_4 \, \partial_1 \partial_2 G(\nu_{21}) \, 
\partial_3 \partial_4 G(\nu_{43})  \nl
& + & \! \! \ve_1 \cdot \ve_3  \, 
\ve_2 \cdot \ve_4 \, \partial_1 \partial_3 G(\nu_{31}) \,
\partial_2 \partial_4 G(\nu_{42})  \label{4int}  \\
& + & \! \! \ve_1 \cdot \ve_4  \, 
\ve_3 \cdot \ve_2 \, \partial_1 \partial_4 G(\nu_{41}) \,
\partial_3 \partial_2 G( \nu_{32}) \, \Big]~.
\nonumber
\eeqa
Using again \eq{newmtgreen}, we can perform the integrals over the punctures,
and we get
\beqa
& & I^{(1)}_4(\tau) = \frac{(2 \pi)^2}{\tau} \sum_{n=1}^{\infty} 
\left(\frac{1 + q^{2n}}{1 - q^{2n}} \right)^2  \nl
& & \times \left[\ve_1 \cdot \ve_3  \,
\ve_2 \cdot \ve_4 - \frac{1}{2}
\ve_1 \cdot \ve_2  \, \ve_3 \cdot \ve_4  \right. \nl
& & \left. - \frac{1}{2} \ve_2 \cdot \ve_3  \, \ve_1 \cdot \ve_4 \right]~.
\label{4res}
\eeqa
The relevant part of the amplitude is then of the form
\beqa
& & \! \! \! \! A^{(1)}(p_1, p_2, p_3, p_4) = \frac{N}{4} 
\frac{g_d^2}{(4 \pi)^{d/2}} (2 \alpha ')^{2 - d/2} \nl 
& \times & Z(d) A^{(0)}(p_1, p_2, p_3, p_4) \, + \, O(\a')~~~,
\label{4onetree}
\eeqa
where the 1PI part of the four-gluon amplitude at tree level in the background
Feynman gauge is given by
\beqa
& & \! \! \! \! \! \! A^{(0)}(p_1, p_2, p_3, p_4) = 4 \, g_{d}^{2} \,  
{\rm Tr} ( \lambda^{a_1} \lambda^{a_2} \lambda^{a_3} \lambda^{a_4} ) \nl
& \times &
\left[ \e_1 \cdot \e_3 \, 
\e_2 \cdot \e_4  - \frac{1}{2} \e_1 \cdot \e_2 
\e_3 \cdot \e_4 \right. \nl
& & \left. - \frac{1}{2} \e_2 \cdot \e_3 \,
\e_1 \cdot \e_4 \right]~.
\label{fourone}
\eeqa

Defining the factor  
\beq
K(d) = \frac{N}{4} \,
\frac{g_d^2}{(4 \pi)^{d/2}} \, (2 \a')^{2 -d/2} \,  Z(d)~~~,
\label{loopefflag}
\eeq
we can now perform the limit $\a' \to 0$, keeping the ultraviolet 
cutoff $\e \equiv 2 - d/2$ small but positive, and eliminating by hand 
the tachyon contribution. The calculation of the integral $Z(d)$ in this
limit can be found in Ref.~\cite{paper}. The result is 
\beq
K(4 - 2 \e) \rightarrow
- \frac{11}{3} \, N \, \frac{g^2}{(4\pi)^2} \, \frac{1}{\e} \,
+ \, O(\e^0)~~~.
\label{divefflag}
\eeq

If we finally compare Eqs. (\ref{renprop1}) with Eqs. (\ref{2onetree}), 
(\ref{3onetree}) and (\ref{4onetree}) we can determine
the renormalization constants. They are given by
\beq
Z_A = Z_3 = Z_4 = 1   + \frac{11}{3} \, N \, \frac{g^2}{(4 \pi)^2} \, 
\frac{1}{\e}~~~,
\label{wardagain}
\eeq
in agreement with the result of the previous section for $Z_A$, and as 
dictated by the background field method Ward identities.

\section{Open string in an external non abelian background}

\label{bosback}

In this section we will study the interaction of an open bosonic string in
an external non-abelian gauge field. In particular we will show that, after 
the integration over the string coordinate, the leading term of the planar
contribution in the small $\alpha' $ expansion reproduces, as expected, the 
usual gauge invariant Yang-Mills
action at each order of string perturbation theory. At one loop,
we can also explicitly evaluate its coefficient, reproducing the wave function 
renormalization constant $Z_A$, as well as \eq{wardagain}.
We see that the connection with the background field method is very general,
and in fact the gluon amplitudes can be understood as interactions of the
string with a particular kind of background, constructed out of plane
waves with  definite momenta.

Let us consider the planar contribution to the partition function of an open 
bosonic string interacting with an external non-abelian $SU(N)$ background.
It is given by 
\beqa
& & \! \! \! \! \! \! \! \! Z_{P}(A_{\mu}) = 
\sum_{h=0}^{\infty} N^h g_s^{2 h -2} 
\int DX Dg ~{\rm e}^{- S_0(X,g;h)} \nl
& & \! \! \! \! \! \! \! \! {\rm Tr} \left[ P_z \exp \left({\rm i} g_d  
\int_h dz ~\partial_z X^{\mu} (z) A_{\mu} ( X (z))    
\right) \right]. \nl
& &  
\label{part}
\eeqa
The path-ordering $P_z$ reminds us that in the open string the $z$ variables
are ordered, as in \eq{cyclord}, along the world-sheet boundary; it
is defined by
\beqa
& & \! \! \! \! \! \! \! \! {\rm Tr} \left[ P_z \exp \left({\rm i} g_d  
\int_h dz ~\partial_z X^{\mu} (z) A_{\mu} ( X (z))    
\right) \right] \nl
& = & \sum_{n=0}^{\infty} \left( {\rm i} g_d  \right)^n \label{pathordered}
\int_{\Gamma_{h,n}} \prod_{i=1}^{n} d z_i  \\
& \times & \partial_{z_1} X^{\mu_1} (z_1 )
\dots \partial_{z_n} X^{\mu_n} (z_n ) \nl  
& \times & {\rm Tr} \left[ A_{\mu_1} (X(z_1 )) \dots A_{\mu_n} 
(X(z_n ))\right]~~~.
\nonumber 
\eeqa 
The precise region of integration for the punctures $z_i$ will in general 
depend on the moduli of the open Riemann surface, and we denoted it by 
$\Gamma_{h,n}$, for a surface of genus $h$ with $n$ punctures. The gauge 
coupling constant $g_d$ and the dimensionless string coupling $g_s$ are 
related by \eq{gstring}. Finally the bosonic string action on a genus $h$ 
manifold with world sheet metric $g_{\alpha \beta}$ is  
\beqa
S_0(X,g;h) & = & \frac{1}{4 \pi \a'} 
\int_h d^2 z ~\sqrt{g} g^{\alpha \beta} \nl 
& \times & \partial_{\alpha} X (z) ~\cdot  ~\partial_{\beta} X(z)~~.
\label{classact}
\eeqa
It is  convenient to separate the zero mode $x^{\mu}$ in the string 
coordinate $X^{\mu}$ from the non zero modes $\xi^{\mu}$ through the relation:
\beq
X^\mu(z) = x^\mu +  (2 \a' )^{1/2} \xi^\mu(z)~~,
\label{xmu}
\eeq
so that $\xi^\mu$ is dimensionless, while the zero mode $x^\mu$ as well as the
string coordinate $X^{\mu}$ have dimensions of length. In terms of the two 
variables $x^{\mu}$ and $\xi^{\mu}$ the measure of the functional integral in 
Eq. (\ref{part}) becomes:
\beq
D X = \frac{d^d x}{(2 \a' )^{d/2}} D \xi
\label{measu}
\eeq 

When we insert in Eq. (\ref{part}) the representation of the path 
ordered exponential given in Eq. (\ref{pathordered})  and we restrict 
ourselves to
the terms up to the order $O(2 \a')^2$ we can write the partition function
in Eq. (\ref{part}) as follows:
\beqa
& & \! \! \! \! \! \! \! \!  Z_{P}(A_{\mu}) = \sum_{h=0}^{\infty} 
\left(\frac{N}{(2 \pi)^d}\right)^h 
g_{s}^{2h-2} \int \frac{d^d x}{(2 \a ')^{d/2}} \nl
& \times & \int d {\cal{M}}_h \Big\{{\rm Tr} (1) - g_{d}^{2} 
\Big[ C_{2}^{(h)} (A) \nl
& & + ~~{\rm i} g_d C_{3}^{(h)} (A)  + ({\rm i} g_d )^2  C_{4}^{(h)} (A)
\Big] \Big\}
\label{expansion}
\eeqa
where $C_i^{(h)}(A)$ is the contribution of the terms 
containing $i$ external gauge fields, and is obtained 
performing the functional integral over the variable $\xi$. The measure of
integration over the moduli $d {\cal{M}}_h$ is equal to the one given in Eq.
(\ref{hmeasure}), but does not include neither the differentials of the 
punctures nor the factor $d V_{abc}$.
For $h > 1$ it  includes also the factor $dV_{abc}$ provided that we do not
fix any of the punctures. In the case of the tree and one loop diagrams there
are not enough moduli to be fixed and therefore another procedure has to be 
followed, as discussed later.

The calculation of the various terms has been discussed in detail in
Ref.~\cite{paper} and will not be reproduced here. Omitting the
vacuum contribution and higher orders in $\alpha '$, we get, for $h >1$
\beqa
Z^{(h)}_{P} ( A_{\mu} ) & = & (2 \a ')^{2 -d/2}~N^h g_{s}^{2h-2}
S_{h} (d) \nl
& \times & \int d^d x \left[- \frac{1}{4}
F_{\mu \nu}^a(x) F_{\mu \nu}^a(x) \right]~,
\label{partfin}
\eeqa
where  
\beqa
S_h (d) & = & - g_{d}^{2} \int d {\cal{M}}_h 
\int dz_1 \int d z_2 \, \theta (z_1 - z_2 ) \nl
& & \! \! \! \! \partial_{z_1}
{\cal G}^{(h)}(z_1 , z_2 )  \partial_{z_2} {\cal G}^{(h)}(z_1 , z_2 )~~,
\label{esse}
\eeqa
and
\beq
F_{\mu \nu}^{a} = \partial_{\mu} A_{\nu}^{a} - \partial_{\nu} A_{\mu}^{a}
+ g_d f^{abc} A_{\mu}^{b} A_{\nu}^{c}~~~.
\label{fmunu}
\eeq

As mentioned before, for $h < 2$ we have to follow another procedure since
there are not enough moduli to be fixed and we prefer not to fix any of the 
punctures in order not to interfere with the definition of path ordering given
by \eq{pathordered} and not to significantly complicate the following 
derivation. However, if we do not fix any of the punctures, we have the problem
that at tree and  one loop loop level the expressions we write are formally 
infinite, because we failed to divide by the volume of the projective group. 
The infinities can be, however, regularized by compactifying the range of 
integration over the punctures, as discussed in Ref.~\cite{tselett} (see also
Ref.~\cite{paper}). Here we give only the results of the calculations.

At tree level, the projective infinity can be regularized by compactifying the 
integration region of the variables $z_i$, mapping them from the real axis 
to a circle. On a circle, following Ref.~\cite{tselett}, we can use the 
Green function 
\beqa
\hat G( \phi_1 , \phi_2 ) & = & \log \left[ 2{\rm i} \sin \left( 
\frac{\phi_1 - \phi_2 }{2} \right)\right] \label{treegreen'} \\
& = & - \sum_{n=1}^{\infty} \frac{\cos n ( \phi_1 - \phi_2 )}{n}
+ \dots~~. \nonumber
\eeqa
The integrals over the punctures $\phi_i $ are now ordered in the interval 
$( 0, 2\pi )$. The dots in Eq. (\ref{treegreen'}) stand for terms 
independent of the punctures, that are irrelevant.

Using \eq{treegreen'}, we find that the basic integral appearing in $S_0(d)$
is given by
\beqa
&  & \! \! \! \! \! \! \! \! \int_{0}^{2 \pi} d \phi_1 \int_{0}^{\phi_1}  
d \phi_2 \,\, \partial_{\phi_1} {\hat G}(\phi_1, \phi_2 ) 
\partial_{\phi_2} {\hat G}
(\phi_1, \phi_2 ) \nl
& = & - \frac{2 \pi }{2} \int_{0}^{2 \pi} d \phi \sum_{n=1}^{\infty} 
\sin^2 n \phi \nl
& = & - \frac{(2 \pi)^2}{4} \sum_{n=1}^{\infty} 1 = 
\frac{(2 \pi)^2}{8}~~,
\label{c2tse}
\eeqa
where we have regularized the sum using $\zeta$-function 
regularization~\cite{mets}.

The previous result implies that $C_2 (A)$ at tree level is equal 
to~\cite{tselett,paper}
\beq 
C_2^{(0)}(A) = - \frac{1}{4} {\tilde{F}}^{a}_{\mu \nu} (x) 
{\tilde{F}}^{a}_{\mu \nu}(x) (2 \a' )^2 \frac{(2 \pi)^2}{8}~.
\label{c2tree}
\eeq
where $ {\tilde{F}}^{a}_{\mu \nu} = \partial_{\mu} A_{\nu}^{a} - 
\partial_{\nu} A_{\mu}^{a}$.

Similarly we can also compute $C_3^{(0)}(A)$ and $C_4^{(0)}(A)$ in 
\eq{expansion} 
and we can see that the full non-abelian gauge invariant action is 
correctly reconstructed, in agreement with 
the results of Ref.~\cite{tselett}, where only $C_2^{(0)}(A)$ is explicitly 
computed while $C_3^{(0)}(A)$ and $C_4^{(0)}(A)$ can be similarly computed 
without fixing any of the punctures. 

The coefficients $C_2^{(1)}(A)$, $C_3^{(1)}(A)$ and $C_4^{(1)}(A)$ can finally
also  be 
computed in the case of one loop obtaining the following one-loop contribution
to the partition function defined by 
\eq{part} 
\beqa
& & \! \! \! \! Z_{P}^{(h=1)} ( A_{\mu} )   =  \frac{N}{4} 
\frac{ g_{d}^{2}}{ (4 \pi)^{d/2}} (2 \a ')^{2 - d/2}\, Z(d) \nl
& \times & \int d^d x  \left[ - \frac{1}{4} F_{\mu \nu}^{a} (x)
F_{\mu \nu}^{a} (x) \right]~,
\label{finh=1}
\eeqa
which is precisely the result of Eq. (\ref{loopefflag}),
with $Z(d)$ given in Eq. (\ref{Zint}).
We have thus verified that the general formalism developed in the first part 
of this section applies to the somewhat special cases $h = 0$ and $h = 1$.

\sect{Concluding remarks}
\label{concl}

We have shown that it is possible to calculate renormalization constants in
Yang-Mills theories using the simplest of string theories, the open bosonic
string. This has been done using a variety of methods, and 
the results concide with the ones obtained using the background field method
and dimensional regularization. Since bosonic string amplitudes are
well understood at all orders in perturbation theory, this technique
may be useful beyond one loop.

\end{document}